\documentclass[aps,prd,superscriptaddress,showpacs,preprint,amsmath,amssymb]{revtex4}
\usepackage{graphicx, bm}
\usepackage{float}
\usepackage{slashed}
\usepackage[usenames]{color}

\begin{document}

\draft

\title{The future muon collider for the research of the anomalous neutral quartic $Z\gamma\gamma\gamma$, $ZZ\gamma\gamma$, and $ZZZ\gamma$ couplings}

\author{ A. Guti\'errez-Rodr\'{\i}guez\footnote{alexgu@fisica.uaz.edu.mx}}
\affiliation{\small Facultad de F\'{\i}sica, Universidad Aut\'onoma de Zacatecas\\
         Apartado Postal C-580, 98060 Zacatecas, M\'exico.\\}

\author{V. Cetinkaya\footnote{volkan.cetinkaya@dpu.edu.tr}}
\affiliation{\small Department of Physics, Kutahya Dumlupinar University, 43100, Turkiye.\\}

\author{M. K\"{o}ksal\footnote{mkoksal@cumhuriyet.edu.tr}}
\affiliation{\small Department of Physics, Sivas Cumhuriyet University, 58140, Sivas, Turkey.}

\author{E. Gurkanli\footnote{egurkanli@sinop.edu.tr}}
\affiliation{\small Department of Physics, Sinop University, Turkey.\\}

\author{V. Ari\footnote{vari@science.ankara.edu.tr}}
\affiliation{\small Department of Physics, Ankara University, Turkey.\\}

\author{ M. A. Hern\'andez-Ru\'{\i}z\footnote{mahernan@uaz.edu.mx}}
\affiliation{\small Unidad Acad\'emica de Ciencias Qu\'{\i}micas, Universidad Aut\'onoma de Zacatecas\\
         Apartado Postal C-585, 98060 Zacatecas, M\'exico.\\}

\date{\today}

\begin{abstract}

In the post-LHC era, the muon collider represents a frontier project capable of providing high-energy and high-luminosity leptonic collisions among future lepton-lepton particle accelerators. In addition, it provides significantly cleaner final states than those produced in hadron
collisions. With this expectation in mind, in this article, we research the sensitivity of the anomalous neutral gauge boson couplings $Z\gamma\gamma\gamma$, $ZZ\gamma \gamma$, and $ZZZ\gamma$ defined by dimension-8 operators, through the $\mu^+\mu^- \to \mu^+\mu^-Z\gamma$ signal, with the $Z$-boson decaying to a neutrino pair. The projections at the future muon collider with a center-of-mass energy of $\sqrt{s}=10$ TeV, integrated luminosity of ${\cal L}=10$ $\rm ab^{-1}$, and systematic uncertainties of $\delta_{sys}=0\%, 3\%, 5\%$, give an expected sensitivity on the anomalous $f_ {T,j}/\Lambda^4$ couplings at $95\%$ confidence level of the order of ${\cal O}(10^{-3}- 10^{-1})$ $\rm TeV^{-4}$. Compared with the research of the ATLAS and CMS Collaborations on the anomalous quartic gauge couplings, we find that the high-luminosity future muon collider could have better sensitivity.

\end{abstract}


\vspace{5mm}

\maketitle


\section{Introduction}

In recent years, the interest of the high-energy scientific community in the design and construction of a possible future muon collider
(MuCol) with multi-TeV energies has been resumed and intensified \cite{Stratakis-2022,Black-2022,Accettura-2023}. In this regard, among
the post-Large Hadron Collider (LHC) generation of particle accelerators, the MuCol represents a frontier project with the capability
to provide very high-energy leptonic collisions and open the path to physics programs beyond the Standard Model (BSM).

The MuCol offers the possibility of collision of particles to very high energies with relatively small circular colliders. Different 
stages of energies of 3, 6, 10, 14 TeV, etc., and very high luminosity are being proposed. In addition, due to their clean collider
environment, this type of collider provides a great tool to search for new physics in the electroweak sector, especially by producing 
multiple electroweak vector bosons. The Standard Model (SM) Lagrangian contains interaction terms involving three or four electroweak 
gauge bosons $(\gamma, Z, W)$. These correspond to the Triple Gauge Couplings (TGC) and Quartic Gauge Couplings (QGC). Studying TGC 
and QGC is essential to test the SM and find evidence of new physics. The SM predicts QGC that contain at least two charged $W$ bosons 
$(W^+W^-W^+W^-, W^+W^-ZZ, W^+W^-\gamma\gamma, W^+W^-Z\gamma)$, while the neutral gauge boson couplings $(ZZZZ, ZZZ\gamma, ZZ\gamma\gamma,  
Z\gamma\gamma\gamma, \gamma\gamma\gamma\gamma)$ are excluded and they are considered to be effects of physics BSM \cite{Data2022}.

One popular approach to studying the anomalous quartic gauge couplings (AQGC) is the concept of Effective Field Theory (EFT), where to parameterize
BSM physics, one can extend the SM Lagrangian by adding higher dimension operators, as is explained in Section II, as well as in Refs. \cite{NPB989-2023,JPG49-2022,EPJC81-210-2021,JPG47-2020,EPJP135-2020,emre-2023}. In this paper, we study the $\mu^+\mu^- \to \mu^+\mu^-Z\gamma \to \mu^- \mu^{+} \nu \bar{\nu} \gamma$ scattering process at a future muon collider sensitive to AQGC. The relevant effective interactions are encoded by dimension-8 operators that lead to the Feynman diagrams shown in Fig. 1.

The most stringent experimental limits arise from vector-boson-scattering processes at the LHC, for which the ATLAS and CMS \cite{CMS:2019qfk,CMS:2020ypo,CMS:2020fqz,JHEP06(2024),EPJC84(2024)} have determined limits of $[-0.12, 0.11]$, $[-0.12, 0.11]$, $[-0.28, 0.28]$,
$[-0.50, 0.50]$, $[-0.40, 0.40]$, $[-0.90, 0.90]$, $[-0.43, 0.43]$, $[-0.92, 0.92]$ TeV$^{-4}$, for the operators $f_{T,j}/\Lambda^4$ with
$j=0, 1, 2, 5, 6, 7, 8, 9$. Limits have also been set by these collaborations using the $pp \to Z\gamma\gamma \to l^+l^-\gamma\gamma$ process 
\cite{JHEP10-2021,PRD93-2016,JHEP10-2017}, and by the CMS and TOTEM Collaborations using the $pp \to pZZp$ process \cite{CMS-TOTEM-2022,Roland-2022}. Previously reported limits include those by the L3 and OPAL Collaborations at the Large Electron Positron collider \cite{PLB540-2002,PRD70-2004}, 
and by the D0 and CDF Collaborations at the Tevatron at Fermilab \cite{PRD62-2000,PRD88-2013}. In summary, AQGCs have been studied at previous, 
present, and proposed hadron-hadron, hadron-lepton, and lepton-lepton colliders \cite{Data2022,Eboli-PRD93-2016,Marantis-JPCS2105-2021,Degrande-AP335-2013,Hao-PRD104-2021,Degrande-JHEP02-2014,Hays-JHEP02-2019,Gounaris-PRD65-2002,
Wrishik-JHEP08-2022,Murphy-JHEP10-2020,Ellis-China64-2021,Murphy-JHEP04-2021,Green-RMP89-2017,Han-PLB1998,Daniele-RNC1997,Boos-PRD1998,Boos-PRD2000,
Beyer-EPJC2006,Christian-EPJC2017,Stephen-arxiv:9505252,Stirling-PLB1999,Belanger-PLB1992,Senol-AHEP2017,Barger-PRD1995,Cuypers-PLB1995,Cuypers-IJMPA1996,
JHEP07-2022}.

In the particular case of the anomalous quartic neutral interactions some studied processes are $e^+e^- \to Z\gamma\gamma$ \cite{Stirling,PRD89-2014}, 
$e^+e^- \to ZZ\gamma$ \cite{EPJC13-2000}, $e^+e^- \to qq\gamma\gamma$ \cite{PLB515-2001}, $e^-\gamma \to ZZe^-$ \cite{Eboli}, $e^-\gamma \to Z\gamma e^-$ \cite{PRD75-2007}, $\gamma\gamma \to ZZ$ \cite{AHEP2016-2016}, $\gamma\gamma \to \gamma Z$ \cite{JHEP10-121-2021}, $\gamma\gamma \to \gamma \gamma$ \cite{EPJC81-2021}, $e^+e^- \to ZZ\gamma$ \cite{EPJP130-2015}, $pp \to Z\gamma jj$ \cite{PRD104-2021}, $pp \to p\gamma\gamma p \to pZ\gamma p$ \cite{JHEP06-142-2017}, $p\bar p \to Z\gamma\gamma$ \cite{JPG26-2000}, $pp(\bar p) \to \gamma\gamma ll$ \cite{Eboli2}, $pp \to p\gamma^*\gamma^* p \to pZZ p$ \cite{PRD81-2010,PRD85-2012,arXiv:0908.2020}, $pp \to p\gamma^* p \to p ZZqX$ \cite{arXiv:1311.1370}, $pp \to p\gamma^*p \to p\gamma ZqX$
\cite{PRD86-2012}, $pp \to qq\gamma ll$ \cite{Eboli4}, and $pp \to Z\gamma\gamma$ \cite{arXiv:2109.12572}. In the case of the future MuCol
the reader can consult Refs. \cite{Inan-2306.03653,Spor-NPB991-2023,Senol-EPJP137-2022,Yang-JHEP09-2022,Yang-JHEP07-2022,Dong-2304.0150,Abbout-2021}.

In the following, after introducing our theoretical framework for the AQGC (Sec. II), we present our strategy for testing for AQGC at the future
MuCol in Sec. III, and finally, we offer our conclusion in Sec. IV.

\section{Effective Field Theory for Non-standard interactions of gauge bosons}

In this section, we briefly collect the various standard and non-standard interactions implied by the effective Lagrangian (1),
as in our previous papers \cite{NPB989-2023,JPG49-2022,EPJC81-2021,JPG47-2020,emre-2023,EPJP135-2020}, as well as in other papers where
the formalism of the AQGC has been widely discussed in the literature \cite{Eboli1,Eboli2,Eboli4,Eboli,Degrande,Eboli3,Eboli-PRD101-2020}.
The EFT extends the SM by adding the higher-dimensional operators that are gauge invariant under the SM:

\begin{equation}
{\cal L}_{eff}= {\cal L}_{SM} +\sum_{k=0}^1\frac{f_{S, k}}{\Lambda^4}O_{S, k} +\sum_{i=0}^{7}\frac{f_{M, i}}{\Lambda^4}O_{M, i}+\sum_{j=0,1,2,5,6,7,8,9}^{}\frac{f_{T, j}}{\Lambda^4}O_{T, j},
\end{equation}

\noindent where each $O_{S, k}$, $O_{M, i}$, and $O_{T, j}$ is a gauge-invariant operator of dimension-8, while $\frac{f_{S, k}}{\Lambda^4}$,
$\frac{f_{M, i}}{\Lambda^4}$, and $\frac{f_{T, j}}{\Lambda^4}$ are the anomalous effective coefficients. A virtue of the EFT formulation
is that one can obtain reliable bounds on the effective coefficients $\frac{f_{S, k}}{\Lambda^4}$, $\frac{f_{M, i}}{\Lambda^4}$, and
$\frac{f_{T, j}}{\Lambda^4}$. These bounds are obtained from general considerations. According to the operator structure, there are three
classes of genuine AQGC operators, as shown in Eq. (3) in Ref. \cite{Degrande}.

Having defined our effective Lagrangian in Eq. (1), we can more precisely state the dimension-8 operators that contain contributions
of operators with covariant derivatives, contributions of operators with gauge field strength tensors, and contributions of operators
with both covariant derivatives and field strength tensors, as follows:

$\bullet$ Operators with the Higgs boson field ($\frac{f_{S, k}}{\Lambda^4} O_{S, k}$):

\begin{eqnarray}
O_{S, 0}&=&[(D_\mu\Phi)^\dagger (D_\nu\Phi)]\times [(D^\mu\Phi)^\dagger (D^\nu\Phi)],  \\
O_{S, 1}&=&[(D_\mu\Phi)^\dagger (D^\mu\Phi)]\times [(D_\nu\Phi)^\dagger (D^\nu\Phi)],  \\
O_{S, 2}&=&[(D_\mu\Phi)^\dagger (D_\nu\Phi)]\times [(D^\nu\Phi)^\dagger (D^\mu\Phi)].
\end{eqnarray}

$\bullet$ Operators with Higgs and gauge boson fields ($\frac{f_{M, i}}{\Lambda^4}O_{M, i}$):

\begin{eqnarray}
O_{M, 0}&=&Tr[W_{\mu\nu} W^{\mu\nu}]\times [(D_\beta\Phi)^\dagger (D^\beta\Phi)],  \\
O_{M, 1}&=&Tr[W_{\mu\nu} W^{\nu\beta}]\times [(D_\beta\Phi)^\dagger (D^\mu\Phi)],  \\
O_{M, 2}&=&[B_{\mu\nu} B^{\mu\nu}]\times [(D_\beta\Phi)^\dagger (D^\beta\Phi)],  \\
O_{M, 3}&=&[B_{\mu\nu} B^{\nu\beta}]\times [(D_\beta\Phi)^\dagger (D^\mu\Phi)],  \\
O_{M, 4}&=&[(D_\mu\Phi)^\dagger W_{\beta\nu} (D^\mu\Phi)]\times B^{\beta\nu},  \\
O_{M, 5}&=&[(D_\mu\Phi)^\dagger W_{\beta\nu} (D^\nu\Phi)]\times B^{\beta\mu} + h.c. ,  \\
O_{M, 7}&=&[(D_\mu\Phi)^\dagger W_{\beta\nu} W^{\beta\mu} (D^\nu\Phi)].
\end{eqnarray}

$\bullet$ Operators with gauge boson fields ($\frac{f_{T, j}}{\Lambda^4} O_{T, j}$):

\begin{eqnarray}
O_{T, 0}&=&Tr[W_{\mu\nu} W^{\mu\nu}]\times Tr[W_{\alpha\beta}W^{\alpha\beta}],  \\
O_{T, 1}&=&Tr[W_{\alpha\nu} W^{\mu\beta}]\times Tr[W_{\mu\beta}W^{\alpha\nu}],  \\
O_{T, 2}&=&Tr[W_{\alpha\mu} W^{\mu\beta}]\times Tr[W_{\beta\nu}W^{\nu\alpha}],  \\
O_{T, 5}&=&Tr[W_{\mu\nu} W^{\mu\nu}]\times B_{\alpha\beta}B^{\alpha\beta},  \\
O_{T, 6}&=&Tr[W_{\alpha\nu} W^{\mu\beta}]\times B_{\mu\beta}B^{\alpha\nu},  \\
O_{T, 7}&=&Tr[W_{\alpha\mu} W^{\mu\beta}]\times B_{\beta\nu}B^{\nu\alpha},  \\
O_{T, 8}&=&B_{\mu\nu} B^{\mu\nu}B_{\alpha\beta}B^{\alpha\beta},  \\
O_{T, 9}&=&B_{\alpha\mu} B^{\mu\beta}B_{\beta\nu}B^{\nu\alpha}.
\end{eqnarray}

It is worth mentioning that the subscripts $S$, $T$, and $M$ in the operators given by Eqs. (2)-(19) refer to scalar (or longitudinal), transverse,
and mixed. These correspond to covariant derivatives of the Higgs field for the longitudinal part and field strength
tensors for the transverse part. In the set of genuine AQGC operators given in Eqs. (2)-(19), $\Phi$ stands for the Higgs doublet, the covariant
derivatives of the Higgs field are given by $D_\mu\Phi=(\partial_\mu + igW^j_\mu \frac{\sigma^j}{2}+ \frac{i}{2}g'B_\mu )\Phi$, where $\sigma^j (j=1,2,3)$
represent the Pauli matrices, and $W^{\mu\nu}$ and $B^{\mu\nu}$ are the gauge field strength tensors for the gauge groups $SU(2)_L$ and $U(1)_Y$.

In this paper, we study the operators that lead to AQGC without an ATGC counterpart. The AQGC interactions caused by each operator are shown 
in Table I. The $\mu^+\mu^- \to \mu^+\mu^-Z\gamma$ signal is especially sensitive to the anomalous coefficients $f_{T,j}/\Lambda^4$, $j=0, 1, 
2, 5, 6, 7, 8, 9$. For this reason, we consider only these operators in our study. The sensitivities obtained for the $f_{T,8}/\Lambda^4$ 
and $f_{T,9}/\Lambda^4$ couplings are of particular interest because they can be extracted only by studying the production of electroweak 
neutral bosons. The contributions of the $O_{T,0}$ and $O_{T,1}$ operators to the examined process are the same \cite{PRD104-2021}, so we
only explicitly study one of the corresponding couplings.

\begin{table}[H]
\centering
\caption{The QGC modified by the listed dimension-8 operators are shown with $\checkmark$.}
\label{tab2}
\begin{tabular}{lccccccccc}
\hline
\hline
Operators & $WWWW$ & $WWZZ$ & $ZZZZ$ & $WW\gamma Z$ & $WW\gamma \gamma$ & $ZZZ\gamma$ & $ZZ\gamma \gamma$ & $Z \gamma\gamma\gamma$ & $\gamma\gamma\gamma\gamma$ \\
\hline
$O_{S0}$, $O_{S1}$, $O_{S2}$        & $\checkmark$ & $\checkmark$ & $\checkmark$ &              &              &              &              &              &              \\
$O_{M0}$, $O_{M1}$, $O_{M7}$        & $\checkmark$ & $\checkmark$ & $\checkmark$ & $\checkmark$ & $\checkmark$ & $\checkmark$ & $\checkmark$ &              &              \\
$O_{M2}$, $O_{M3}$, $O_{M4}$, $O_{M5}$ &             & $\checkmark$ & $\checkmark$ & $\checkmark$ & $\checkmark$ & $\checkmark$ & $\checkmark$ &              &              \\
$O_{T0}$, $O_{T1}$, $O_{T2}$        & $\checkmark$ & $\checkmark$ & $\checkmark$ & $\checkmark$ & $\checkmark$ & $\checkmark$ & $\checkmark$ & $\checkmark$ & $\checkmark$ \\
$O_{T5}$, $O_{T6}$, $O_{T7}$        &             & $\checkmark$ & $\checkmark$ & $\checkmark$ & $\checkmark$ & $\checkmark$ & $\checkmark$ & $\checkmark$ & $\checkmark$ \\
$O_{T8}$, $O_{T9}$                  &             &              & $\checkmark$ &              &              & $\checkmark$ & $\checkmark$ & $\checkmark$ & $\checkmark$ \\
\hline
\end{tabular}
\end{table}

\section{$\mu^+\mu^-Z\gamma$ production and sensitivities on anomalous couplings}

\subsection{Number of events and total cross-section}

\begin{figure}[!h]
\centerline{\scalebox{1.05}{\includegraphics{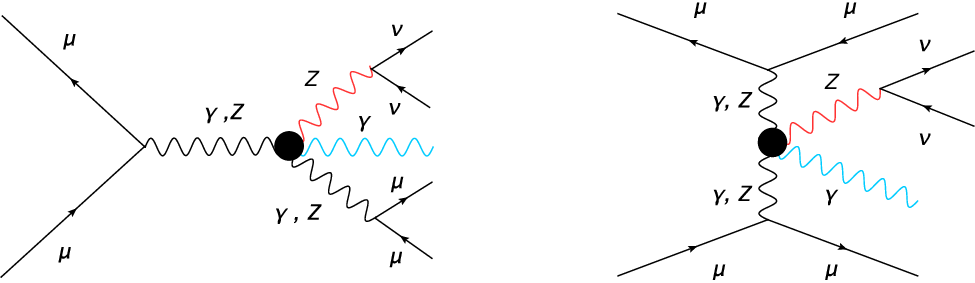}}}
\caption{Feynman diagrams of the process $\mu^- \mu^{+} \to \mu^- \mu^{+} Z\gamma $ including the anomalous
contribution of $ZZ\gamma \gamma$, $Z\gamma\gamma \gamma$, and $ZZZ\gamma$ vertices. New physics (represented by a black circle) in the electroweak
sector can induce the neutral AQGC.}
\label{Fig.1}
\end{figure}

We focus on the possibility of direct measurement of the total cross-section of the associated production channel $\mu^+\mu^- \to \mu^+\mu^-Z\gamma$,
with anomalous couplings $f_{T,j}/\Lambda^4$, $j=0-2, 5-9$ at a MuCol, and in the framework of the EFT. The Feynman diagrams with the $Z\gamma\gamma\gamma$, $ZZ\gamma\gamma$, and $ZZZ\gamma$ vertices are shown
in Fig.~\ref{Fig.1}

The total cross-section of the reaction $\mu^+\mu^- \to \mu^+\mu^-Z\gamma\to \mu^- \mu^{+} \nu \bar{\nu} \gamma$ contains the following contributions:

\begin{eqnarray}
\sigma_{Tot}\Biggl(\sqrt{s}, \frac{f_{T,j}}{\Lambda^{4}}\Biggr)
&=& \sigma_{SM}( \sqrt{s} ) +\sigma_{(Int)}\Biggl( \sqrt{s}, \frac{f_{T,j}}{\Lambda^{4}}\Biggr)
+ \sigma_{NP}\Biggl(\sqrt{s}, \frac{f^2_{T,j}}{\Lambda^{8}} \Biggr), \hspace{3mm} j= 0-2, 5-9,    \nonumber\\
\end{eqnarray}

\noindent where $\sigma_{SM}$ is the SM, $\sigma_{Int}$ is the interference term between the SM and the EFT operators, and $\sigma_{NP}$ is the
contribution of new physics.

\begin{figure}[!h]
\centerline{\scalebox{1.3}{\includegraphics{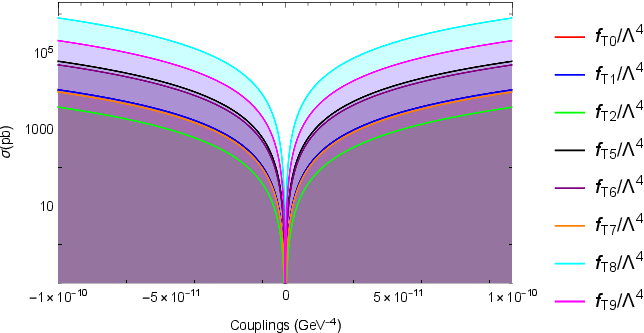}}}
\caption{Total cross-section for the process $\mu^+\mu^- \to \mu^+ \mu^- Z\gamma$ with the $Z$-boson
decaying to a neutrino pair, and in terms of the anomalous $f_{T,0}/\Lambda^4$, $f_{T,1}/\Lambda^4$,
$f_{T,2}/\Lambda^4$, $f_{T,5}/\Lambda^4$, $f_{T,6}/\Lambda^4$, $f_{T,7}/\Lambda^4$, $f_{T,8}/\Lambda^4$,
and $f_{T,9}/\Lambda^4$ couplings for the muon collider with $\sqrt{s}=10$ TeV. This figure is for purely
illustrative purposes. For a perturbative coupling $f<1$ the EFT is valid for effective coupling values $< 10^{-16}$ ${\rm GeV^{-4}}$.}
\label{Fig.5}
\end{figure}

In Fig.~\ref{Fig.5}, the process $\mu^+\mu^- \to \mu^+\mu^-Z\gamma$ with the $Z$-boson decaying to a neutrino pair at $\sqrt{s}=10$ TeV is analyzed by individually varying the parameters $f_{T,j}/\Lambda^4$, while keeping the remaining coefficients fixed at zero to isolate their contributions. When all coefficients are zero, the cross-section aligns with the SM prediction ($\sigma_{SM}$). This methodology facilitates the computation of total cross-sections for varying coupling, revealing deviations from the SM. The findings illustrate the muon collider's capability to probe these anomalous couplings, establishing its effectiveness in exploring aQGCs within the EW sector.

The numerical calculation has been implemented with the help of the MadGraph5\_aMC@NLO \cite{MadGraph} code, where the operators given
in Eqs. (12)-(19) are implemented into MadGraph5\_aMC@NLO through the FeynRules package \cite{AAlloul} as a Universal FeynRules Output (UFO)
module \cite{CDegrande}. We have used the requirements on the rapidity, missing transverse energy, and transverse momentum of the outgoing photons, $|\eta^{\gamma}|< 2.5$ (Cut-1), $\slashed{E}_T > 1000$ GeV (Cut-2), $p^\gamma_T > 800$ GeV (Cut-3) given in Table II to suppress the background 
process $\mu^- \mu^{+} \to \mu^- \mu^{+} \nu \bar{\nu} \gamma $.

\begin{table}[!h]
\centering
\caption{Particle-level selection cuts for the $\mu^{+} \mu^{-} Z\gamma$ signal at the muon collider.}
\label{tab1}
\begin{tabular}{cc}
\hline
\hline
Kinematical Cuts & \qquad Anomalous Couplings $f_{T,j}/\Lambda^4$ for j =0,1,2,5,6,7,8 and 9  \\
\hline
Cut-1   & $|\eta^{\gamma}| < 2.5$ \\
Cut-2   & $\slashed{E}_T > 1000$ GeV \\
Cut-3   & $p^\gamma_T > 800$ GeV \\
\hline

\hline
\end{tabular}
\end{table}

We have used the kinematical distributions of the final state particles with generator-level cuts (minimal cuts applied by MadGraph) to obtain the optimized cuts and reveal the separation region. The distributions are composed of five sets of the anomalous couplings $f_{T,0, 2, 5, 7, 8}/\Lambda^4$+SM as signal
and the non-resonant process $\mu^- \mu^{+} \to \mu^- \mu^{+} \nu \bar{\nu} \gamma $ as background. Here, the SM refers to the cross-section of the process $\mu^+\mu^- \to \mu^+\mu^-Z\gamma\to \mu^- \mu^{+} \nu \bar{\nu} \gamma $ fixing all EFT couplings to zero.

\begin{figure}[H]
\centerline{\scalebox{0.90}{\includegraphics{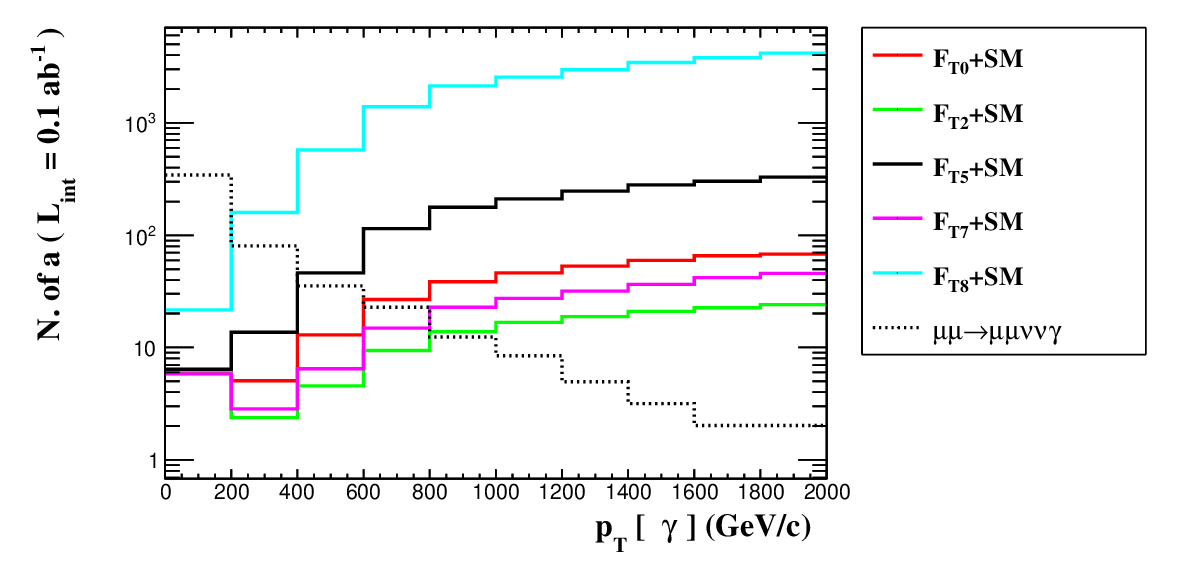}}}
\caption{ \label{fig:gamma} The number of expected events as a function of photon transverse momentum $p^{\gamma}_T$
for the $\mu^- \mu^{+} \to \mu^- \mu^{+} Z\gamma \to \mu^- \mu^{+} (\nu \bar{\nu}) \gamma $ signal and backgrounds at
the MuCol with $\sqrt{s}=10$ TeV. The distributions are for $f_ {T,0,2,5,7,8}/\Lambda^4$ and relevant background.
In this figure, we have taken a value of  $0.1\hspace{1mm}{\rm TeV}^{-4}$ for each anomalous coupling. This value
is chosen for illustrative purposes and is well above that which would correspond to a perturbative coupling $f=1$
and an energy scale larger than the center-of-mass energy of the collider (10 TeV).}
\label{Fig.3}
\end{figure}

Fig.~\ref{Fig.3} illustrates the behavior of the number of events of the $\mu^+\mu^- \to \mu^+\mu^-Z\gamma \to \mu^- \mu^{+} \nu \bar{\nu} \gamma$ 
signal and of the background process, as a function of the photon transverse momentum $p^{\gamma}_T$. From the figure, it can be seen that for 
$p^{\gamma}_T > 800$ GeV, the majority of the signal remains while the majority of background is removed.

\begin{figure}[H]
\centerline{\scalebox{0.90}{\includegraphics{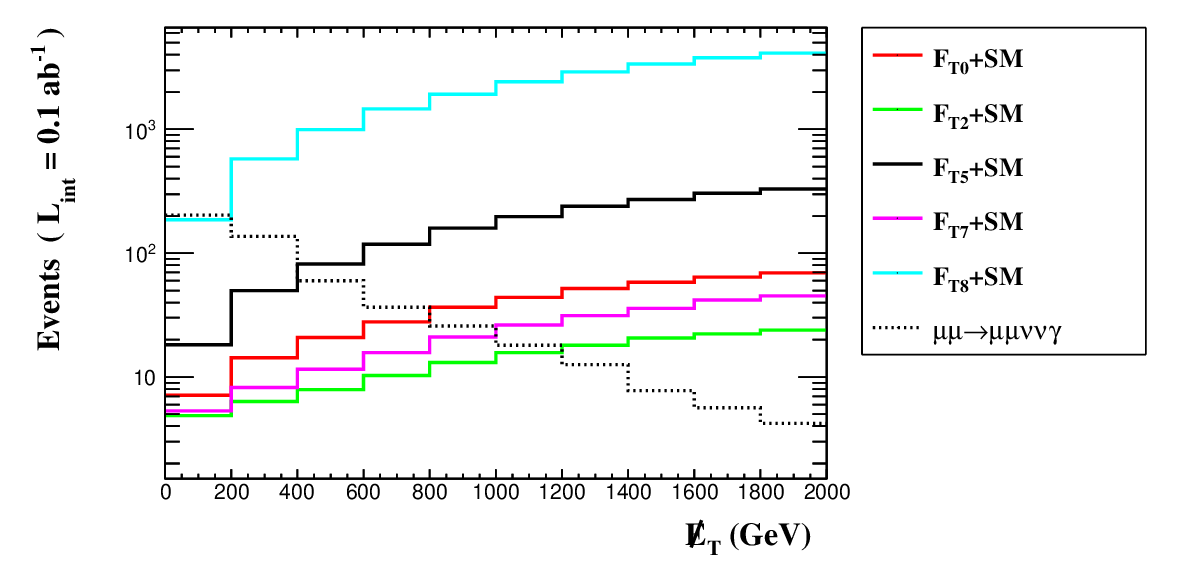}}}
\caption{The number of expected events as a function of the missing transverse energy $\slashed{E}_T$ (GeV)
for the $\mu^- \mu^{+} \to \mu^- \mu^{+} Z\gamma \to \mu^- \mu^{+} (\nu \bar{\nu}) \gamma $ signal with $f_ {T,0,2,5,7,8}/\Lambda^4$ and
relevant backgrounds at the MuCol. In this figure, we have taken a value of  $0.1\hspace{1mm}{\rm TeV}^{-4}$ for each anomalous coupling.
This value is chosen for illustrative purposes and is well above that which would correspond to a perturbative coupling $f=1$ and an energy
scale larger than the center-of-mass energy of the collider (10 TeV).}
\label{Fig.4}
\end{figure}

Fig.~\ref{Fig.4} shows the photon $\slashed{E}_T$ distributions for the predicted signal from $\mu^+\mu^-Z\gamma$ AQGC for $f_{T,0, 2, 5, 7, 8}/\Lambda^4$
and relevant background at the MuCol. The figure shows that the majority of signal has $\slashed{E}_T > 1000$ GeV while the majority of background has $\slashed{E}_T$ below this value.

\subsection{Sensitivity of the anomalous $Z\gamma\gamma\gamma$, $ZZ\gamma \gamma$, and $ZZZ\gamma$ vertices at the future muon collider}

In this subsection, we evaluate the sensitivity limits at $95\%$ C.L. of the eight Wilson coefficients $f_{T,j}/\Lambda^4$, $j=0, 1, 2, 5, 6, 7, 8, 9$
through the $\mu^+\mu^- \to \mu^+ \mu^- Z\gamma \to \mu^- \mu^{+} \nu \bar{\nu} \gamma$ signal, and for the energy and luminosity of a possible future
muon collider with $\sqrt{s}=10$ TeV, ${\cal L}=10$ $\rm ab^{-1}$. A simple and practical method to estimate the sensitivity of the $f_{T,j}/\Lambda^4$ parameters is based on the chi-square distribution:

\begin{equation}
\chi^2(f_{T,j}/\Lambda^4)=\Biggl(\frac{\sigma_{BSM}(\sqrt{s}, f_{T,j}/\Lambda^4)-\sigma_{SM}(\sqrt{s})-\sigma_{BG}(\sqrt{s})}
{[\sigma_{SM}(\sqrt{s})+\sigma_{BG}(\sqrt{s})]\sqrt{(\delta_{st})^2 + (\delta_{sys})^2}}\Biggr)^2,
\end{equation}

\noindent where $\sigma_{SM}(\sqrt{s})$ is $\mu^+\mu^- \to \mu^+ \mu^- Z\gamma$ and $\sigma_{BG}(\sqrt{s})$ is the non-resonant $\mu^+\mu^- \to \mu^- \mu^{+}
\nu \bar{\nu} \gamma$ process, respectively. $\sigma_{BSM}(\sqrt{s}, f_{T,j}/\Lambda^4)$ represents the cross-section in the presence of BSM interactions, $\delta_{st}=\frac{1}{\sqrt{N_{SM}}}$ is the statistical error and $\delta_{sys}$ is the relative systematic uncertainty. The primary sources
of systematic uncertainty stem from the cross-section predictions of both signal and background processes. Other factors contributing to uncertainty include integrated luminosity, efficiency in identifying photons, and uncertainties associated with the energy-momentum scales and resolutions of final-state particles. While a detailed discussion of systematic uncertainty sources is not the primary focus of this study, our goal is to explore the overall impact of systematic uncertainty on the sensitivities of AQGC. Therefore, we examine three scenarios of systematic uncertainty to assess its effects comprehensively. We choose $\delta_{sys}=0\%, 3\%$, and $5\%$ for our numerical analysis. The number of events is given by $N_{SM}={\cal L}\times \sigma_{SM}$, where ${\cal L}$ is the integrated luminosity of the MuCol.

Table III lists the Wilson coefficients $f_{T,j}/\Lambda^4$, $j=0, 1, 2, 5, 6, 7, 8, 9$ and their dependence on the assumed
systematic uncertainty. From this table, the anomalous couplings $f_{T,5}/\Lambda^4$, $f_{T,6}/\Lambda^4$, $f_{T,8}/\Lambda^4$, and $f_{T,9}/
\Lambda^4$ show a greater sensitivity, of the order of ${\cal O}(10^{-3}-10^{-2})$ TeV$^{-4}$, compared to the others remaining in the set.
In addition, our results show up to 2-3 orders of magnitude better sensitivity compared to the experimental
results reported by the ATLAS and CMS Collaborations at the LHC \cite{JHEP10-2021,PRD93-2016,JHEP10-2017,CMS:2019qfk,CMS:2020ypo,CMS:2020fqz,JHEP06(2024),EPJC84(2024)}, as well as many other experimental and phenomenological results. The obtained results on AQGC in the study are competitive with the results reported in papers \cite{Inan-2306.03653,Spor-NPB991-2023,Senol-EPJP137-2022,Yang-JHEP09-2022,Yang-JHEP07-2022,Dong-2304.0150,Abbout-2021}.

\begin{figure}[H]
\centerline{\scalebox{1.5}{\includegraphics{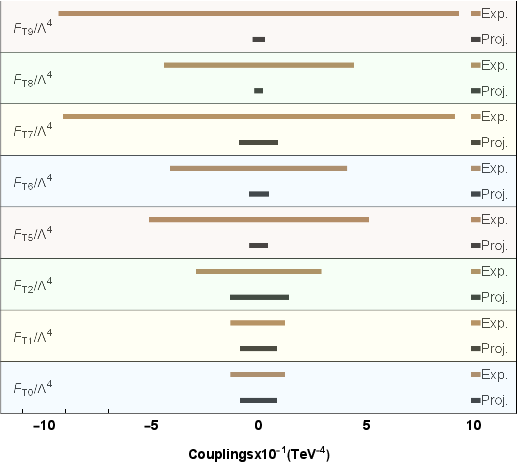}}}
\caption{Comparison of the current experimental limits \cite{CMS:2019qfk,CMS:2020ypo,CMS:2020fqz,JHEP06(2024),EPJC84(2024)}
and projected sensitivity on the anomalous $f_{T,0}/\Lambda^4$, $f_{T,1}/\Lambda^4$, $f_{T,2}/\Lambda^4$, $f_{T,5}/\Lambda^4$,
$f_{T,6}/\Lambda^4$, $f_{T,7}/\Lambda^4$, $f_{T,8}/\Lambda^4$, and $f_{T,9}/\Lambda^4$ couplings for expected luminosity
of ${\cal L}=10\hspace{0.8mm}$ ab$^{-1}$ with $\sqrt{s}=10\hspace{0.8mm}{\rm TeV}$ at the muon collider. Here, the x-axis
illustrates the \textbf{Couplings $\times 10^{-1}$}. For a perturbative value of $f (<1)$, the valid EFT region is $<10^{-4}$
${\rm TeV}^{-4}$ for a 10 TeV muon collider.}
\label{Fig.7}
\end{figure}

Finally, Fig.~\ref{Fig.7} illustrates the comparison of the most sensitive experimental limits and projected sensitivity on the anomalous $f_{T,0}/\Lambda^4$, $f_{T,1}/\Lambda^4$, $f_{T,2}/\Lambda^4$, $f_{T,5}/\Lambda^4$, $f_{T,6}/\Lambda^4$, $f_{T,7}/\Lambda^4$, $f_{T,8}/\Lambda^4$, and $f_{T,9}/\Lambda^4$
couplings for the center-of-mass energy and integrated luminosity expected of $\sqrt{s}=10\hspace{0.8mm}{\rm TeV}$ and ${\cal L}=10\hspace{0.8mm}$ ab$^{-1}$,
at the future MuCol. The figure shows a significant difference between the experimental results \cite{CMS:2019qfk,CMS:2020ypo,CMS:2020fqz,JHEP06(2024),EPJC84(2024)} and the projected sensitivities. These results indicate that our study improves the existing analyses on the AQGC of the proposed present and future hadron-hadron, hadron-lepton, and lepton-lepton colliders, and provides further guidance for the viable processes in the search for new physics through the anomalous neutral $Z\gamma\gamma\gamma$, $ZZ\gamma\gamma$, and $ZZZ\gamma$ couplings that can be investigated in a possible future muon collider at the frontiers of energy and luminosity.

\begin{table}[H]
\centering
\caption{Expected sensitivities at $95\%$ C.L. to the anomalous $ZZ\gamma\gamma$, $Z\gamma\gamma\gamma$, and $ZZZ\gamma$ couplings through the process $\mu^+\mu^- \to \mu^+ \mu^- Z \gamma \to \mu^- \mu^{+} \nu \bar{\nu} \gamma$ at the muon collider, for systematic uncertainties of $\delta_{sys}=0\%, 3\%, 5\%$, center-of-mass energy of 10 TeV and the integrated luminosity of ${\cal L}=10$ $\rm ab^{-1}$. For a perturbative value of $f (<1)$, the valid EFT region is $<10^{-4}$ ${\rm TeV}^{-4}$ for a 10 TeV muon collider.}
\label{tab3}
\begin{tabular}{ccccc}
\hline
\hline
\multicolumn{2}{c}{Couplings (TeV$^{-4}$)} & Experimental Results (ATLAS \& CMS) & Systematic Errors & Our Projection \\

\hline\hline
                       &  &  & $\delta=0\%$ & $[-7.62,7.36]\times10^{-2}$  \\
$f_{T0}/\Lambda^{4}$&  & $[-1.20,1.10]\times10^{-1}$ \cite{CMS:2019qfk} & $\delta=3\%$ & $[-7.81,7.55]\times10^{-2}$  \\
                       &  &  & $\delta=5\%$ & $[-7.95,7.69]\times10^{-2}$  \\
\hline
                       &  &  & $\delta=0\%$ & $[-7.62,7.36]\times10^{-2}$  \\
$f_{T1}/\Lambda^{4}$   &  & $[-1.20,1.10]\times10^{-1}$ \cite{CMS:2019qfk} & $\delta=3\%$ & $[-7.81,7.55]\times10^{-2}$  \\
                       &  &  & $\delta=5\%$ & $[-7.95,7.69]\times10^{-2}$  \\
\hline
                       &  &  & $\delta=0\%$ & $[-1.23,1.29]\times10^{-1}$  \\
$f_{T2}/\Lambda^{4}$   &  & $[-2.80,2.80]\times10^{-1}$ \cite{CMS:2019qfk} & $\delta=3\%$ & $[-1.27,1.32]\times10^{-1}$  \\
                       &  &  & $\delta=5\%$ & $[-1.29,1.34]\times10^{-1}$  \\
\hline
                       &  &  & $\delta=0\%$ & $[-3.24,3.07]\times10^{-2}$  \\
$f_{T5}/\Lambda^{4}$   &  & $[-5.00,5.00]\times10^{-1}$ \cite{CMS:2020ypo} & $\delta=3\%$ & $[-3.32,3.15]\times10^{-2}$  \\
                       &  &  & $\delta=5\%$ & $[-3.37,3.21]\times10^{-2}$  \\
\hline
                       &  &  & $\delta=0\%$ & $[-3.43,3.53]\times10^{-2}$  \\
$f_{T6}/\Lambda^{4}$   &  & $[-4.00,4.00]\times10^{-1}$ \cite{CMS:2020ypo} & $\delta=3\%$ & $[-3.52,3.62]\times10^{-2}$  \\
                       &  &  & $\delta=5\%$ & $[-3.58,3.69]\times10^{-2}$  \\
\hline
                       &  &  & $\delta=0\%$ & $[-7.88,7.78]\times10^{-2}$  \\
$f_{T7}/\Lambda^{4}$   &  & $[-9.00,9.00]\times10^{-1}$ \cite{CMS:2020ypo} & $\delta=3\%$ & $[-8.08,7.98]\times10^{-2}$  \\
                       &  &  & $\delta=5\%$ & $[-8.22,8.13]\times10^{-2}$  \\
\hline
                       &  &  & $\delta=0\%$ & $[-8.56,8.57]\times10^{-3}$  \\
$f_{T8}/\Lambda^{4}$   &  & $[-4.30,4.30]\times10^{-1}$ \cite{CMS:2020fqz} & $\delta=3\%$ & $[-8.77,8.78]\times10^{-3}$  \\
                       &  &  & $\delta=5\%$ & $[-8.93,8.94]\times10^{-3}$  \\
\hline
                       &  &  & $\delta=0\%$ & $[-1.71,1.69]\times10^{-2}$  \\
$f_{T9}/\Lambda^{4}$   &  & $[-9.20,9.20]\times10^{-1}$ \cite{CMS:2020fqz} & $\delta=3\%$ & $[-1.75,1.73]\times10^{-2}$  \\
                       &  &  & $\delta=5\%$ & $[-1.78,1.76]\times10^{-2}$  \\
\hline
\end{tabular}
\end{table}

\section{CONCLUSION}

In this paper, we have evaluated the cross-section for the $\mu^+\mu^- \to \mu^+ \mu^- Z \gamma \to \mu^- \mu^{+} \nu \bar{\nu} \gamma$ process
in $\mu\mu$ collisions at a center-of-mass energy of 10 TeV corresponding to an integrated luminosity of 10 $\rm ab^{-1}$. In addition, we choose
three values for the systematic uncertainty of $\delta_{sys}=0\%, 3\%,$ and $5\%$.

The total cross-sections are calculated in a fiducial region where simulated signal events are selected at generator level in the $\mu^+ \mu^- Z
\gamma$ channel by requiring exactly two muons and at least one photon with $p^\gamma_T > 800$ GeV. The photon candidate must have pseudorapidity $|\eta^\gamma| < 2.5$, and events must satisfy $\slashed{E}_T > 1000$ GeV.

With the elements mentioned above, we estimate projected sensitivities on the anomalous quartic gauge couplings $f_{Tj}/\Lambda^{4}$ using the
$\mu^+\mu^- \to \mu^+ \mu^- Z \gamma$ channel at the center-of-mass energy of $\sqrt{s}=10$ TeV and integrated luminosity of ${\cal L}=10$ ab$^{-1}$
for a possible future MuCol. The results reported in this paper are given in Fig.~\ref{Fig.7} and Table III, respectively.

In this article, we have shown the importance of the future high-energy and high-luminosity muon collider for the search of deviations from the SM
via the anomalous $Z\gamma\gamma\gamma$, $ZZ\gamma\gamma$, and $ZZZ\gamma$ vertices encoded in the Wilson coefficients $f_{Tj}/\Lambda^4$. For collider
energy equal to 10 TeV and $\delta=0\%$, a sensitivity to the Wilson coefficients of $f_{T8}/\Lambda^{4}=[-8.56,8.57]\times10^{-3}$ TeV$^{-4}$, $f_{T9}/\Lambda^{4}=[-1.71,1.69]\times10^{-2}$ TeV$^{-4}$, and $f_{T5}/\Lambda^{4}=[-3.24,3.07]\times10^{-2}$ TeV$^{-4}$ can be reached by the future MuCol. This is significantly better than the latest results of the ATLAS and CMS Collaborations, where the sensitivity for $f_{T, j}/\Lambda^4$, with $j=0, 1, 2, 5, 6, 7, 8, 9$ is ${\cal O}(10^{-1})$ \cite{CMS:2019qfk,CMS:2020ypo,CMS:2020fqz,JHEP06(2024),EPJC84(2024)}. A comparison between the MuCol predictions and the measurements made by the ATLAS and CMS Collaborations for the anomalous $Z\gamma\gamma\gamma$, $ZZ\gamma\gamma$, and $ZZZ\gamma$ couplings shows that the future MuCol could achieve a sensitivity of up to one order of magnitude better than latest results of ATLAS and CMS. In this sense, the paper could be relevant for the scientific community to prioritize future searches and experimental efforts.

Without a doubt, the future MuCol represents a unique project in elementary particle physics that can provide very high-energy and high-luminosity muon collisions and open the path to a vast and primarily unexplored physics program.

\vspace{1.5cm}

\begin{center}
{\bf Acknowledgements}
\end{center}

A. G. R. and M. A. H. R. thank SNII and PROFEXCE (M\'exico). The numerical calculations reported in this paper were fully performed
at TUBITAK ULAKBIM, High Performance and Grid Computing Center (TRUBA resources).

\vspace{1cm}

\noindent {\bf Declarations}  \\
$\bullet$ Data Availability Statement: All data generated or analyzed during this study are included in this article.

\vspace{1cm}


\end{document}